\input harvmac
\def\np#1#2#3{Nucl. Phys. B {#1} (#2) #3}

\def\plb#1#2#3{Phys. Lett. B {#1} (#2) #3}

\def\physrev#1#2#3{Phys. Rev. D {#1} (#2) #3}

\def\dim{{\rm dim}}

\def\mI{{\cal M}_{Inst}}
\def\FI{Fayet-Iliopoulos}
\def\cA{{\cal A}}
\def\V{{\cal V}}

%  draw box of size #1pt and line thickness #2pt
\def\drawbox#1#2{\hrule height#2pt 
        \hbox{\vrule width#2pt height#1pt \kern#1pt 
              \vrule width#2pt}
              \hrule height#2pt}
% Young tableaux

\def\Fund#1#2{\vcenter{\vbox{\drawbox{#1}{#2}}}}
\def\Asym#1#2{\vcenter{\vbox{\drawbox{#1}{#2}
              \kern-#2pt       % line up boxes
              \drawbox{#1}{#2}}}}
 
\def\fund{\Fund{6.5}{0.4}}
\def\asym{\Asym{6.5}{0.4}}
\batchmode
  \font\bbbfont=msbm10
\errorstopmode
\newif\ifamsf\amsftrue
\ifx\bbbfont\nullfont
  \amsffalse
\fi
\ifamsf
\def\IR{\hbox{\bbbfont R}}
\def\IC{\hbox{\bbbfont C}}

\def\IZ{\hbox{\bbbfont Z}}
\def\IF{\hbox{\bbbfont F}}
\def\IP{\hbox{\bbbfont P}}
\else
\def\IR{\relax{\rm I\kern-.18em R}}
\def\IZ{\relax\ifmmode\hbox{Z\kern-.4em Z}\else{Z\kern-.4em Z}\fi}
\def\IF{\relax{\rm I\kern-.18em F}}
\def\IP{\relax{\rm I\kern-.18em P}}
\fi
\nref\jptasi{J. Polchinksi, ``TASI Lectures on D-Branes,''
hep-th/9611050.}
\nref\wsmall{E. Witten, 
hep-th/9511030, \np{460}{1995}{541.}}
\nref\GP{E. G. Gimon and Polchinski,  hep-th/9601038,
\physrev{54}{1996}{1667}.}
\nref\dm{M. Douglas and G. Moore,  hep-th/9603167.}
\nref\kn{P. B. Kronheimer and H. Nakajima,  Math. Ann. 288 (1990) 263.}
\lref\mdegs{M.R. Douglas,  hep-th/9612126.}
\nref\obrane{K. Intriligator,  hep-th/9702038, Nucl. Phys. B to
appear.}
\nref\ABPSS{C. Angelantonj, M. Bianchi, G. Pradisi, A. Sagnotti,
Ya. S. Stanev, hep-th/9607229, \plb{387}{1996}{743}.}
\nref\BiSa{M. Bianchi and A. Sagnotti, \plb{247}{1990}{517}; 
\np{361}{1991}{519}.} 
\nref\DabPari{A. Dabholkar and J. Park, 
 hep-th/9602030, \np{472}{1996}{207}.}
\nref\GJ{E. G. Gimon and C. V. Johnson, 
hep-th/9604129, \np{477}{1996}{715}.}
\nref\DabParii{A. Dabholkar and J. Park, 
hep-th/9604178, \np{477}{1996}{701}.}
\nref\pol{J. Polchinski, 
hep-th/9606165 .}
\nref\GJtwo{E. G. Gimon and C. V. Johnson,  
hep-th/9606176, \np{479}{1996}{285}.}
\nref\BluZaff{J. D. Blum and A. Zaffaroni,  \plb{387}{1996}{71}.}
\nref\GopMuk{R. Gopakumar and S. Mukhi,  \np{479}{1996}{260}.}
\nref\Julie{J. D. Blum,  hep-th/9608053, \np{486}{1997}{34.}}
\lref\aspin{P. S. Aspinwall,  hep-th/9612108.}
\lref\mrdbranes{M. Douglas, hep-th/9512077.}
\lref\aspinB{P.S. Aspinwall,  hep-th/9507012, \plb{357}{1995}{329}.}
\lref\Betal{M. Berkooz, R.G. Leigh, J. Polchinski, J. Schwarz,
N. Seiberg, and E. Witten,  hep-th/9605184, \np{475}{1996}{115}.}
\lref\kron{P.B. Kronheimer, Jour. Differential Geometry, 
{\bf 29} (1989) 665.}
\lref\bi{J. D. Blum and K. Intriligator, to appear.}
\lref\CS{E. Cremmer and J. Scherk,  \np{72}{1974}{117}.}
\lref\wbound{E. Witten, 
hep-th/9510135, \np{460}{1996}{335}.}
\lref\sw{N. Seiberg and E. Witten, hep-th/9603003, \np{471}{1996}{121}.}
\lref\wcomments{E. Witten, hep-th/9507121, Proc. of Strings '95.}
\lref\sopen{A. Strominger, hep-th/9512059, \plb{383}{1996}{44}.}
\lref\witfb{E. Witten, hep-th/9512219, \np{463}{1996}{383}.}
\lref\horwit{P. Horava and E. Witten, hep-th/9510209,
\np{460}{1996}{506}.}
\Title{hep-th/9705030, IASSNS-HEP-97/39}
{\vbox{\centerline{Consistency Conditions for Branes}
\centerline{at Orbifold Singularities}}}
\medskip
\centerline{Julie D. Blum and 
Kenneth Intriligator\footnote{${}^*$}{On leave 1996-1997
{}from Department of Physics, University of California, San Diego.}}
\vglue .5cm
\centerline{School of Natural Sciences}
\centerline{Institute for Advanced Study}
\centerline{Princeton, NJ 08540, USA}

\bigskip
\noindent

We discuss consistency conditions for branes at orbifold
singularities.   The conditions have a world-sheet interpretation in
terms of tadpole cancellation and a space-time interpretation in terms
of anomalies.  As examples, we consider type II and type I branes
on $\IC ^2/\IZ _M$ orbifolds.  We give orientifold constructions of
phases of type I or heterotic string theory, involving branches
with extra tensor multiplets, which arise when small $SO(32)$
instantons sit on orbifold singularities.

\Date{5/97}                                   
%\draftmode   

\newsec{Introduction}

The familiar miracle of string theory is that phenomena on the
world-sheet carry over to space-time.  In particular, consistency of
the theory on the world-sheet implies that the space-time theory is
free of anomalies.    Our interest here will be relating world-sheet
consistency conditions to space-time consistency conditions for
quantum field theories living inside D branes.  (See 
\jptasi\ for a recent review with references on D branes.)
The QFTs living in the world-volume of D branes can be considered
without having to include gravity and other stringy modes by taking
the $M_p\rightarrow \infty$ limit.  Though the world-volume QFTs may
need more data, such as string theory, to obtain a sensible theory in
the ultra-violet, at long distances they must be sensible on their
own.  In particular, as gauge anomalies only depend on the massless
spectrum, the world-volume QFTs must be free of gauge anomalies.

For example, in \wsmall\ it was argued that an $SO(32)$ heterotic or
type I instanton of zero size, which is a D5 brane, has a
supersymmetric (with 8 super-charges) $Sp(1)\cong SU(2)$ gauge theory
with $N_f=16$ fundamental flavors living in its world-volume.  With
$K$ such D5 branes on top of each other, the 6d world-volume gauge
theory becomes $Sp(K)$ with $N_f=16$ matter fields in the $\fund $ and
one in the $\asym$.  When the $SO(32)$ heterotic or type I string
theory is compactified to six dimensions on $K3$, there is a
restriction that $K\leq 24$ since the total number instantons, large
and small, must be 24 in order to have zero $H$ charge on the compact
$K3$.  But for the uncompactified string theories, the number $K$ of
D5 branes is completely arbitrary -- their $H$ flux can go off to
infinity.  Therefore, for any $K$, the above $Sp(K)$ gauge theory must
be free of $6d$ gauge anomalies; this is, indeed, true.  It would not
have been true for $SO(N)$ with $N\neq 32$.

World-sheet tadpole consistency conditions were discussed in \GP\ in
the context of compactification of type I string theory on a $K3$
realized as a $T^4/\IZ _2$ orientifold.  These conditions ensure that
the theory living in the uncompactified six dimensions is free of all
gauge and gravitational anomalies.  In \dm, type II and type I string
theory was considered on non-compact $\IR ^4/\IZ _M$ orbifolds and
orientifolds.  It was pointed out there that the tadpole consistency
conditions can be ignored in the non-compact context because such
conditions give, for example, a total charge $\int dH$ which only
needs to vanish on a compact space.  It was also pointed out in \dm\
that this is not completely satisfactory as, for example, the type I
anomaly condition which requires $SO(32)$ on $\IR ^{10}$ comes from a
tadpole condition which is not a source for any physical field.

More generally, on physical grounds, some consistency conditions can
be ignored in non-compact cases, corresponding to the fact that
fluxes and gravity can leak out into the non-compact dimensions.
Other consistency conditions, on the other hand, must apply even in
the non-compact context.  For example, even in the non-compact
context, there must be consistency conditions which ensure that the
QFTs living in the world-volume of D branes are anomaly free.  These
anomalies can not flow out of the branes into the non-compact
dimensions.

Our main point is that this distinction between space-time conditions
which can and cannot be ignored in the non-compact context is
perfectly matched by world-sheet considerations.  Keeping the
dependence on the compactification volume $\V$, certain tadpoles are
inversely proportional to $\V$ and thus automatically vanish, without
imposing any conditions, in the $\V\rightarrow \infty$ limit.
Precisely these tadpoles are associated with space-time conditions
which no longer need apply when flux or gravity can leak off into the
non-compact dimensions.  Other tadpoles, on the other hand, do not
automatically vanish in the $\V\rightarrow \infty$ limit and thus can
not be ignored in the non-compact context.  These tadpoles give
precisely the space-time anomaly conditions which must be satisfied
even in the non-compact context.

As examples, we extend the analysis of
\dm\ of branes on (possibly blown up)
orbifold singularities $\IC ^2/\IZ _M$ by pointing out certain tadpole
consistency conditions and their relation to anomaly conditions in the
world volume of the branes. Our consistency relations will be for $p$
and $p+4$ branes on the orbifold spaces.  In the type IIA context, for
example, we can take $p=0,2,4$; for type IIB, $p=1,3$; for type
I, $p=5$.

In the next section, we discuss the world-sheet tadpole
consistency conditions and their relation to space-time conditions.
In sect. 3, we consider type II on the $\IZ _M$ orbifold singularity.
We complete the observation of \dm\ that the hyper-Kahler quotient
theories of \kn\ arise physically in the world-volume of branes at the
orbifold singularity by showing that tadpole consistency conditions
give a relation of \kn\ between the numbers of colors and flavors of
the gauge theory.  By noting the mixing between a $U(1)$ field
strength and the NSNS $B$ field, this relation allows us to derive the
value of the $B$ field for the orbifold; we find agreement with
results obtained in \mdegs\ by another method.  

In sect. 4 we summarize the space-time analysis of \obrane\ for the 6d
theories living in the world-volume of type I small instantons at $\IC
^2/\IZ _M$ singularities.  In particular, we review the argument,
based on space-time anomalies, that there is a ``Coulomb
branch\foot{In terminology following the discussion in \sw.}'' in
which there are specific gauge theories, along with specific numbers
of tensor multiplets, living in the D5 branes at orbifold
singularities.  In sect. 5 we present the orientifold construction of
type I on the orbifold singularities.  We verify that the tadpole
anomalies which remain relevant, even though $\IC ^2/\IZ _M$ is of
infinite volume, reproduce the space-time anomaly conditions found in
\obrane.  We also verify that the orientifold construction
automatically gives exactly the right number of tensors for cancelling
the reducible part of the gauge anomaly.  Thus orientifolds give a
definite construction of the ``Coulomb branch'' phases of \obrane.

Indeed, we point out that orientifold constructions {\it necessarily}
give phases with extra tensor multiplets\foot{See, however, \ABPSS\
for exotic examples with varying numbers of tensor multiplets,
including one with no tensor multiplets at all -- not even the usual
one from the gravity multiplet, which usually contains the dilaton.}.
The single exception to this is the original example of
\refs{\BiSa,\GP}.  In other words, orientifolds never realize standard
perturbative type I theories at ALE singularities.  Indeed, all other
examples of type I orientifolds contained extra tensors; see, for
example \refs{\BiSa -\Julie}.  The interpretation of the extra
tensor multiplets is briefly discussed in sect. 6.

\newsec{Tadpoles and space-time consistency conditions}

Our results indicate that the irreducible parts of the gravitational
and gauge anomalies of the six dimensional theories on D5 branes are
related to the one loop tadpole anomalies arising from the divergent
part of the Klein bottle, Mobius strip, and cylindrical amplitudes of
the string theory on the orbifold.  We illustrate this relation
in the following sections.  A similar relation was previously found
\Julie\ on compact orbifolds arising from F theory.

Remembering that strings in directions transverse to a brane have
Dirichlet boundary conditions, which allow for winding, while those
tangent to the brane have Neumann boundary conditions, which allow for
momentum, we can determine the volume dependence of the various
tadpoles.  Let $\V _4$ be the volume of the four dimensional orbifold,
in our case $\IC ^2/\IZ _M$, $\V _{p+1}$ be the spacetime volume of
the $p$ brane, and $\V _{5-p}$ the volume transverse to these two
spaces.  There are generally three types of tadpole anomalies \GP: the
untwisted $p+4$ brane anomalies, with dependences on volumes $\V _4 \V
_{p+1}/\V _{5-p}$; the untwisted $p$ brane anomalies, with dependence
on volumes $\V _{p+1}/\V _4\V _{5-p}$; and the twisted p brane
anomalies, with dependence on volumes $\V _{p+1}/\V _{5-p}$.  In the
limit where $\V _4
\rightarrow \infty$, the untwisted $p$-brane anomaly automatically
vanishes and can, thus, be ignored.  This anomaly determines the number
of p branes (small instantons) and will always be a free parameter in
our models.

In the type I case $p=5$, and the untwisted nine-brane anomaly will
always determine the number of nine-branes to be $32$.  There is no
freedom here because flux has nowhere to go since $\V _{5-p}$ is
trivial.  The twisted tadpole conditions precisely ensure that the 6d
gauge theories living in the world-volume of the five-branes are free
of spacetime gauge anomalies.  In fact, as we illustrate in sect. 5,
it is possible to extract (up to constants) the individual irreducible
$\tr F^4$ and the reducible $\tr F^2 \tr F^2$ anomaly terms from the
tadpole terms.

In type IIB theories, anomaly cancellation does not allow for any
nine-branes. However, $p+4$($p$) brane theories with $p<5$ will have
no spacetime anomalies.  The string theory still senses the ten
dimensional origin of the theories in that tadpoles can not be truly
ignored until the dimension of $\V _{5-p}$ is large enough so that all
flux can escape to infinity.  In analogy with electrodynamics, we
would not expect this to occur until $p\leq 2$.  Although it is not
associated with an anomaly in the type IIB case, we can constrain the
theories by first solving the twisted tadpole conditions for $\V
_{5-p}$ finite.  We then assume that, in the limit that all flux can
escape to infinity, these conditions are still valid.  This assumption
is perhaps not entirely justified but is supported by the fact, shown
in the next section, that it gives the correct instanton moduli space.

\newsec{Type II on a $\IC ^2/\IZ _M$ orbifold}

\subsec{General aspects}
$N$ coincident type II $p+4$ branes have a supersymmetric $U(N)$ gauge
theory living in their world-volume \wbound.  $K$ coincident type II
$p$ branes living inside of this system have a supersymmetric (with
eight super-charges) $U(K)$ gauge theory living in their world volume
with $N$ matter hypermultiplets in the ${\bf K}$ and one in the
adjoint ${\bf K^2}$.  This configuration of branes has the
interpretation as $K$ $U(N)$ instantons and, indeed, this $U(K)$
world-volume theory has a Higgs branch which gives the hyper-Kahler
quotient construction of the moduli space of $K$ $U(N)$ instantons
\mrdbranes.

We now consider the $p$ branes on a possibly blown-up singularity
$X\cong \IC ^2/\IZ _M$.  $X$ is hyper-Kahler with a triple of Kahler
forms $\vec \omega$; $X$ has $M-1$ non-trivial two-cycles $\Sigma _i$,
$i=1\dots M-1$, which generate $H_2$, and
\eqn\blowup{\int _{\Sigma _i}\vec \omega =\vec \zeta _i,}
with $\vec \zeta _i$ the $M-1$ blowing up parameters.  Because $\pi
_1(X_{\infty})=\IZ _M$, there can be group elements $\rho _{\infty}\in
U(N)$ at infinity representing $\IZ _M$: $\rho _\infty$ has $w_\mu$
eigenvalues $e^{2\pi i \mu /M}$, $\mu =0\dots M-1$, with $w_\mu n_\mu
=N$.  We define $n_\mu \equiv 1$, the $SU(M)$ Dynkin indices; all
greek indices run from $0\dots M-1$, with repeated indices
summed. $\rho _\infty$ breaks $U(N)\rightarrow
\prod _{\mu =0}^r U(w_\mu )$.  

In addition, because $X$ has non-trivial two cycles $\Sigma _i$,
$i=1\dots M-1$, there can also be non-trivial first Chern classes
$u_i\equiv\int _{\Sigma _i}\tr\ F$.  We note that 
\eqn\uBreln{u_i=N\int _{\Sigma _i}B,}
where $B$ is the usual NSNS two-form $B_{\mu \nu}$ field.  The
relation \uBreln\ follows from the mechanism of
\CS, which was applied to $D$ strings in
\wbound: in the presence of the Dirichlet boundary conditions, the
world sheet $B$ field mixes with the field strength of the gauge field
which couples to the Dirichlet boundary.  The gauge invariant field
strength two-form for the overall $U(1)$ gauge field coupling to the
boundary is ${\cal F}=N^{-1}\tr F-B$, where the $N^{-1}$ accounts for
the trace.  Because the two-cycles $\Sigma _i$ are closed, $\int
_{\Sigma _i}{\cal F}=0$, which gives \uBreln.  It is only
the NSNS $B$ field, which couples to winding, which mixes with the
gauge field strength; the RR $B$ field does not enter in ${\cal F}$.
Thus in type I, where the NSNS $B$ field is projected out, there is no
such mechanism for getting non-trivial first Chern classes.

The moduli space $\mI$ of $U(N)$ instantons on arbitrary ALE spaces
was given a hyper-Kahler quotient construction in \kn; i.e. $\mI$ is
isomorphic to the Higgs branch of a supersymmetric gauge theory with
eight super-charges.  For our case of $\IC ^2/\IZ _M$, the gauge group
is $\prod _{\mu =0}^{M-1}U(v_\mu)$ with matter in the $\oplus _\mu
w_\mu \fund _\mu$ and $\oplus _{\mu} (\fund _\mu , \fund _{\mu +1})$,
where the subscript labels the gauge group and $\fund _M\equiv \fund
_0$.  The numbers of colors, $v_\mu$, are related to the physical data
$w_\mu$ and $u_i$ discussed above via \kn\
\eqn\cvw{\widetilde C_{\mu \nu}v_\nu =w_\mu -u_\mu.}
$\widetilde C_{\mu \nu}$ is the extended Cartan matrix for $SU(M)$ and
$u_0$ is determined by $N\equiv w_\mu n_\mu=u_\mu n_\mu$, which
follows from \cvw\ and $\widetilde C_{\mu \nu}n_\mu =0$.  The solution
of \cvw\ is
\eqn\unvis{v_\mu =K+G_{\mu \nu}(w_\nu -u_\nu ),}
where $K$ is an arbitrary non-negative integer and $G_{\mu \nu}=G_{\nu
\mu}$ is defined by $G_{\mu<\nu }\equiv \mu (M-\nu)/M$.

\subsec{The orbifold calculation}

It was shown in \dm\ that the gauge theories of \kn, reviewed in the
previous subsection, arise physically in the world-volume of type II
branes at the orbifold singularities.  In \dm\ the tadpole equations
were not imposed and the physical restriction \cvw\ on the $v_\mu$ was
not obtained.  As discussed in the previous section, however, the
twisted sector tadpoles should constrain the gauge group.  We show
that they indeed yield \cvw\ with, via \uBreln, a specific value for
the $B$ field.  

Following the discussion in \dm, $\gamma_g$ will represent the action
of the generator, $g$, of $\IZ _M$ on the Chan-Paton factors, thus
\eqn\conthree{\gamma_g^M=1.}
We can always choose a $\gamma_g$ satisfying \conthree\ to be a
diagonal matrix of dimension $w\cdot n$ (dimension $v\cdot n$) for
$p+4$ ($p$) branes such that $w_{\mu}$ ($v_{\mu}$) components along the
diagonal are $\alpha_{\mu}=e^{2\pi i\mu
\over M}$.  As shown in \dm, the orbifold indeed yields a gauge theory
with $\prod _{\mu }U(v_\mu )$ gauge group and matter content discussed
above.  

The twisted tadpoles calculated by \GJ\ are directly applicable for
type II instantons on the $\IZ _M$ orbifold: we simply drop their
cross-cap terms. For every $k$ in $1\leq k\leq M-1$, the following
tadpole equations must then be satisfied to avoid a divergence (for
finite $\V _{5-p}$):
\eqn\gju{{1\over 4\sin^2 
({\pi k\over M})}(\sum_{\mu =0}^{M-1}w_{\mu}e^{{2\pi i k\mu\over
M}}-4\sin^2 ({\pi k\over M})v_{\mu}e^{{2\pi i k\mu\over M}})^2 =0.}  A
small manipulation converts these conditions to exactly the conditions
\cvw, though with a specific value of the first Chern classes: 
$u_{\mu}= Nn_\mu/M$.  It follows from \uBreln\ that
\eqn\orbB{\int _{\Sigma _i}B={1\over M}}
for all $i=1\dots M-1$.  As in \aspinB, in the orbifold construction
of the theory on $\IC ^2/\IZ_M$, there is thus a fixed non-zero value
of the $B$ field turned on.  The result \orbB\ for the value of the
$B$ field was also obtained in \mdegs\ by requiring $D0$ branes to
have the correct mass.

\newsec{Type I or heterotic -- space-time analysis}

We now consider type I or heterotic small $Spin(32)$ instanton D5
branes sitting at the $\IC ^2/\IZ _M$ singularity, reviewing the
space-time analysis of \obrane.  As in the previous section, in
addition to the number $K$ of $D5$ branes, it is necessary to specify
$\rho _\infty\in Spin(32)$ representing $\IZ _M$. Following the
discussion in \Betal, the gauge group in type I or heterotic is
actually $Spin(32)/\IZ _2$, where the $\IZ _2$ is generated by the
element $w$ in the center of $Spin(32)$ which acts as $-1$ on the
vector, $-1$ on the spinor of negative chirality, and $+1$ on the
spinor of positive chirality. Because only representations with $w=1$
are in the $Spin(32)/\IZ _2$ string theory, the identity element $e\in
\IZ _M$ can be mapped to either the element $1$ or $w$ in $Spin (32)$.
Thus either $\rho _\infty ^M=1$, which will be referred to as the case
with possible\foot{The ``possible'' qualifier is because the notion of
vector structure is actually ill-defined on the Coulomb branch
\aspin.}  vector structure, or $\rho _\infty ^M=w$, which will be
referred to as the case without vector structure.

\subsec{Case $\rho _\infty ^M=1$ with possible vector structure}

The group element $\rho _\infty \in Spin(32)$ which satisfies $\rho
_\infty ^M=1$ has $w_\mu$ eigenvalues $e^{2\pi i \mu /M}$, $\mu
=0\dots M-1$, with $w_\mu n_\mu =32$ and $w_\mu =w_{M-\mu}$ in order
to have $\rho _\infty \in Spin(32)$; $\rho _\infty$ breaks
$Spin(32)\rightarrow Spin(w_0)\times U(w_1)\times \cdots \times
U(w_{P-1})\times Spin(w_P)$ for $M=2P$ and $Spin(32)\rightarrow
Spin(w_1)\times U(w_1)\times \cdots \times U(w_P)$ for $M=2P+1$.

The relevant small instanton gauge theory for $M=2P$ is 
\eqn\evengg{Sp(v_0)\times U(v_1)\times U(v_2)\times \cdots \times
U(v_{P-1})\times Sp (v_P), \qquad M=2P, } with hypermultiplets $\half
w_0 \cdot \fund _0$, $\oplus _{j=1}^{P-1}w_j\cdot \fund _j$, $\half
w_P\cdot \fund _P$, and $\oplus _{j=1}^{P}(\fund _{j-1} , \fund _j)$
(subscripts label the gauge group).  For $M=2P+1$ the relevant gauge
theory is
\eqn\oddgg{Sp(v_0)\times U(v_1)\times U(v_2)\times \cdots \times
U(v_{P-1})\times U(v_P), \qquad M=2P+1, } with hypermultiplets in the
$\half w_0 \cdot \fund _0$, $\oplus _{j=1}^{P}w_j\cdot \fund _j$,
$\oplus _{j=1}^P (\fund _{j-1} , \fund _j)$, and $\asym _P$.  The
theories \evengg\ and \oddgg\ are given, respectively, by the ``I.4''
and ``I.2'' quiver diagrams described in sect. 4.4 of \dm.  They have
the unbroken space-time gauge symmetry as global symmetries.  

The theories \evengg\ and \oddgg\ have a non-trivial six dimensional
gauge anomaly:
\eqn\AzMis{\cA =\half \sum _{\mu =0}^{M-1}(
\widetilde C_{\mu \nu}V_\nu -w_\mu +D_\mu)\tr F_\mu ^4+3\sum
_{j=1}^P(\tr F_{j-1}^2 -\tr F_j^2)^2;} $F_\mu$ for $\mu >P$ is defined
by $F_\mu \equiv F_{M-\mu}$, $\widetilde C_{\mu\nu}\equiv 2\delta
_{\mu \nu}-a_{\mu \nu}$ is the Cartan matrix of the extended $SU(M)$
Dynkin diagram, and $D_\mu \equiv 8(2\delta _{\mu, 0}+\delta _{\mu
,P}+\delta _{\mu ,M-P})$.
We define $V_{\mu}$ for $\mu=0
\dots M-1$ by $V_{\mu\leq P}\equiv \dim (\fund _\mu)$ and $V_{\mu
>P}\equiv V_{M-\mu}$; i.e. $V_0\equiv 2v_0$, $V_{i<P}\equiv v_i$, and
$V_P\equiv 2v_P$ ($V_P\equiv v_P$) for $M=2P$ ($M=2P+1$).  Also, as
the theories
\evengg\ and
\oddgg\ only have $M-P-1$ $U(1)$ factors to which \FI\ terms can be
coupled, they are missing $P$ hypermultiplet blowing up moduli.

It was conjectured in \obrane\ that there is a ``Coulomb
branch'' phase of the heterotic or type I string theory where the
theory in the world volume of the D5 branes is of the form \evengg\ or
\oddgg, but with $29P$ hypermultiplets traded for 
$P$ additional tensor multiplets.  In order to cancel the irreducible
$\tr F_\mu ^4$ part of the gauge anomaly \AzMis, it is necessary to
have \eqn\cvwd{\widetilde C_{\mu \nu}V_\nu =w_\mu -D_\mu;} this gives
$w_\mu n_\mu =32$ as expected.  The solution of
\cvwd\ is 
\eqn\VMp{V_{\mu \leq P} =2K+\sum _{\nu =0}^P{\rm min}(\mu ,
\nu)W_\nu -8\mu,} 
with $K$ an arbitrary positive integer (which is large enough so that
all $V_\mu \geq 0$) and $W_{\nu <P}\equiv w_\mu$ and $W_P\equiv w_P$
($W_P\equiv \half w_P$) for $M$ odd (even).  With the relation \cvwd,
there are $28P$ missing hypermultiplets associated with deforming the
instanton moduli; remembering the $P$ missing blowing up moduli
mentioned above, there are $29P$ missing hypermultiplets.  Finally, in
order to cancel the reducible $\tr F_\mu ^2 \tr F_\nu ^2$ anomaly in
\AzMis, it is necessary to introduce $P$ tensor multiplets 
with couplings to the gauge fields in \evengg\ or \oddgg\ which are
the supersymmetric completion of the interactions
\eqn\phiFi{\sum _{i=0}^P(\Phi _{i+1}-\Phi _i)\tr F_i^2,}
where $\Phi _i$ are the real scalar components of the tensor
multiplets, with $\Phi _0\equiv \Phi _{P+1}\equiv 0$.

\subsec{Case $\rho _\infty ^M=w$ without vector structure}

This case requires $M=2P$; $\rho _\infty$ has $w_j$ eigenvalues
$e^{i\pi (2j-1)/2P}$, $j=1\dots 2P$, with $w_{2P+1-j}=w_j$ and $\sum
_{j=1}^{2P}w_j=2\sum _{j=1}^P=32$ for $\rho _{\infty}\in Spin(32)$.

It was conjectured in \obrane\ that the world-volume theory has a
``Coulomb branch'' phase with $29(P-1)$ hypermultiplets traded for $P-1$
tensor multiplets.  The gauge theory on the Coulomb branch is $\prod
_{i=1}^PU(v_i)$ with matter content $\oplus _{i=1}^Pw_i\cdot
\fund _i$, $\oplus _{i=1}^{P-1}(\fund _i, \fund _{i+1})$, $\asym _1$,
and $\asym _P$.  This is described by the ``type I5'' quiver diagrams
of \dm.  The gauge anomaly is 
\eqn\anov{\cA=\half \sum _{i=1}^{2P}(\sum _{j=1}^{2P}
\widetilde C_{ij}v_j-w_i+D_i)\tr F_i^4+3\sum _{r=1}^{P-1}(\tr F_r^2
-\tr F_{r+1}^2)^2,} where $\widetilde C _{ij}$ is the Cartan matrix
for the extended $SU(2P)$ Dynkin diagram, $F_{i>P}\equiv F_{2P+1-i}$,
$v_{i>P}\equiv v_{2P+1-i}$, and $D_i\equiv 8(\delta _{i,1}+\delta
_{i,P}+\delta _{i,2P}+\delta _{i, P+1})$.  Thus, in order to cancel
the irreducible $\tr F_i ^4$ gauge anomaly, it is necessary to have
\eqn\vwnovec{\sum _{j=1}^{2P}\widetilde C_{ij}v_j=w_i-D_i;}
note that this properly gives $w_\mu n_\mu =32$.  The solution of
\vwnovec\ is
\eqn\vnovs{v_{i\leq P}=2K+\sum _{j=1}^P{\rm min}(i-1,j-1)w_j-8(i-1),}
where $K$ is an arbitrary non-negative integer (which is large enough
so that all $v_i\geq 0$).

In order to cancel the
reducible gauge anomaly in \anov, it is necessary to have $P-1$ tensor
multiplets with coupling to the gauge fields $F_i$ of $U(v_i)$ which
are the supersymmetric completion of
\eqn\phiFnov{\sum_{i=1}^P(\Phi _i-\Phi _{i-1})\tr F_i ^2,}
where $\Phi _i$ are the scalar components of the tensor multiplets,
with $\Phi _0\equiv \Phi _P\equiv 0$.  

\newsec{The orientifold analysis}

In this section, we derive via a direct orientifold analysis the
``Coulomb branch'' phases \obrane\ reviewed in the previous section.
As the relevant gauge theories were already obtained in
\dm\ by an orientifold construction, all that remains to do is to: explain
why there are the extra tensor multiplets and missing
hyper-multiplets, obtain the relations \cvwd\ and \vwnovec, and obtain
the couplings \phiFi\ and \phiFnov.

In the following we call $\Omega$ the element that reverses the
orientation of a type IIB string and $\gamma_{\Omega}$ the matrix
representing the action of $\Omega$ on Chan-Paton factors.  In
addition to the condition \conthree, there are two more algebraic
algebraic consistency conditions \refs{\GP , \dm}:
\eqn\conone{\gamma_{\Omega}=\pm\gamma_{\Omega}^T}
where the upper(lower) sign is for 9(5) branes, and 
\eqn\contwo{\gamma_g\gamma_{\Omega}\gamma_g^T=\alpha\gamma_{\Omega}.}
The case without vector structure will always have $\alpha=
\alpha_1$ and can only occur for $M$ even, while the case with 
possible vector structure will have $\alpha=\alpha_0=1$.

The above two conditions imply that for $M$ even there are two
possible actions of $\Omega$ on Chan-Paton factors; they correspond to
with or without possible vector structure.  In the convention of
\GJ, the action of $\Omega$ on the Chan-Paton factors for these
two cases is such that 
\eqn\choice{\tr(\gamma_{\Omega k}^{-1}\gamma_{\Omega k}^{T})=
\epsilon\, \tr(\gamma_{\Omega k+M/2}^{-1}\gamma_{\Omega k+M/2}^{T})}
where $\epsilon=+1(-1)$ for the case with (without) vector structure.
The two possible actions of $\gamma _\Omega$ correspond to two
different actions of $\Omega$ on the world-sheet.  Some of the
following is similar to a discussion of \pol.  There is only one
choice for the action of $\Omega$ in the untwisted sector that will
not project out the graviton.  In each of the other twisted sectors we
have {\it a priori} two choices for $\Omega$, one of which will
preserve a tensor multiplet and the other of which will preserve a
hypermultiplet.  We will show that tree-loop duality of the
non-oriented and open string amplitudes reduces this choice to the
$\IZ _2$ twisted sector.  The case without vector structure will
preserve the tensor while the other case will preserve the
hypermultiplet.  Given the choice of $\Omega$, we fix $\epsilon$ and
the $\IZ _2$ twisted sector of the loop Klein bottle.  This choice, in
turn, fixes the signs of the Mobius strip amplitudes.  The relative
signs of all cross-caps are then fixed by the duality relation.  Since
we can only obtain two possibilities for consistent tadpole
amplitudes, this means that the sign of $\Omega$ is relatively fixed
in the other twisted sectors.  What is the sign of $\Omega$?  Let us
suppose that we have a closed bosonic string state twisted by $g^k$
propagating around a cylinder.  Now apply $\Omega$ and close the
cylinder to create a Klein bottle.  We know this amplitude vanishes by
loop-tree duality for $k\neq 0, M/2$.  This implies that $\Omega$ acts
in one of two ways.  It either switches the $k$ twist to the $M-k$
twist (in other words it includes $J$ \GJtwo), which then vanishes in
the trace, or $\Omega$ acts with a different sign to the left than to
the right so that the amplitude vanishes.  The second choice is
equivalent to $\Omega$ having a relative sign between the $k$ twisted
and $M-k$ twisted sectors for $k\neq 0,M/2$, and we will assume
corresponds to the case with vector structure ($\epsilon =+1$).  $J$
corresponds to the case without vector structure ($\epsilon =-1$).

Both of the above choices for the action of $\Omega$ mean that the
twisted sectors with $k\neq 0, \half M$ yield ${M\over 2}-1$
(${M-1\over 2}$) extra tensors for $M$ even (odd), along with the same
number of hyper-multiplets.  These hyper-multiplets transform, as in
the discussion in \refs{\dm, \Betal}, to cancel the anomalies
associated with the $U(1)$ factors in the gauge groups.  They pair
up with the $U(1)$ gauge fields to give them a mass and their
expectation values become \FI\ parameters.  Because the above argument
for the presence of extra tensor multiplets is quite general,
orientifolds can never realize standard type I theories at ALE
singularities.

The twisted tadpoles exactly reproduce (up to constants) the
individual terms in the space-time anomalies.  Our general procedure
will be as follows.  The twisted tadpoles will be of the form $\sum
_{g \neq 1}(a^g _\mu w_\mu +b_\mu ^g v_\mu +c^g )^2$. To obtain the
irreducible $\tr (F_{\mu})^4$ term in the anomaly, note that this term
in the tadpole should be proportional to $tr(I_{v_{\mu}})=v_{\mu}$
($I_n$ is the $n\times n$ dimensional identity matrix), and this term
will thus be proportional to $\sum _{g\neq 1}b^g _\mu
(a_\nu ^g w_\nu +b_\nu ^g v_\nu +c^g )$.  Similarly, to
obtain the reducible $\tr(F_{\mu})^2
\tr(F_{\nu})^2$ term, we extract the coefficient
of $v_{\mu} v_{\nu}$, which will be $\sum _{g\neq 1}b^g
_\mu b^g _\nu$.  The other pieces of the tadpole are the
reduction of the $p+5$ dimensional gauge anomaly to $p+1$ dimensions
and the gravitational anomaly.  In fact, the irreducible gravitational
anomaly should have the coefficient $\sum _{g\neq 1}c^g
(a^g _\mu w_\mu +b^g _\mu v_\mu +c^g )$.  (For type IIB,
$c^g =0$ so there is no $p+1$ dimensional gravitational anomaly.)

{}From the reducible term, we obtain the coefficient (up to a constant)
of the twisted sector $B$ fields coupling to the gauge fields.
The same $B$ fields also couples to the nine-brane gauge fields with
the coefficient of $w_\mu w_\nu$\foot{When combined with the fact that
$\int _{ALE}dH \neq 0$ with five-branes present, this means that there
is an anomaly inflow mechanism, with net current flowing onto the
branes from the ten-dimensional world.  This is the limit of the
mechanism of
\ref\jbjh{J. Blum and J. Harvey, hep-th/9310035, \np{416}{1994}{119}.}
where the instantons shrink to zero size.}.  The tensor $B$ components
of the tensor multiplet come from the twisted Ramond-Ramond sector
while the real scalar comes from the twisted NS-NS sector; they have
the same couplings, as is required by supersymmetry, since it is the
same tadpole for NS-NS and R-R.

\subsec{Case with possible vector structure}
  
To derive the above results from the orientifold, we augment the
results of \dm\ with those of \GJ.  In the closed string calculation,
we must assume that an extra minus exists in the action of $\Omega$
for sectors twisted by $1-t$ relative to those twisted by $t$ for
$0<t<{1\over 2}$.  The extra minus is also present for $t={1\over 2}$.
In sectors with the extra minus, the three-brane is not projected out,
and one obtains a tensor multiplet rather than a hypermultiplet.  This
extra minus changes the sign of terms twisted by $\IZ _2$ in the one
loop Klein bottle so that the tadpoles of \GJ\ are modified.  Using
the solutions of $\gamma_{\Omega}$ and $\gamma_g$ derived by \dm, we
obtain $w\cdot n=32$ and
\eqn\gjonev{{1\over 4\sin^2 ({\pi k\over M})}(\sum_{\mu =0}^{M-1}
w_{\mu}e^{{2\pi i k\mu\over M}}-4\sin^2 ({\pi k\over
M})v_{\mu}e^{{2\pi i k\mu\over M}}- 32\delta_{k,0\, mod\, 2})^2=0}
for $M$ even, and
\eqn\gjodd{{1\over 4\sin^2 ({2\pi k\over M})}(\sum_{\mu =0}^{M-1}
w_{\mu}e^{{4\pi i
k\mu\over M}}-4\sin^2 ({2\pi k\over M})v_{\mu}e^{{4\pi i
k\mu\over M}}- 32\cos^2({\pi k\over\ M}))^2=0}
for $M$ odd.  Using the method discussed above for extracting the
appropriate terms, these tadpoles yield exactly the space-time anomaly
\AzMis\ (up to the relative factor of $6$ between the reducible and
irreducible parts).

To summarize, the orientifold analysis gives exactly the ``Coulomb
phase'' spectrum and tensor multiplet couplings of \obrane.  For the
$\IZ _2$ case, the Coulomb phase had previously been constructed via
$F$ theory in the context of a compact $K3$ (thus there was an upper
bound on the possible $K$) \aspin.

\subsec{Case without vector structure}

In the orientifold calculation, the extra minus in the $\IZ _2$
twisted sector is no longer present.  Thus, in the $\IZ _2$ twisted
closed string sector, the three-brane is now projected out, and we
obtain a hypermultiplet rather than a tensor, giving one less tensor
as compared to the possible vector structure case; there are then a
total of $P-1$ extra tensor multiplets.  To get the anomalies we use
the $\gamma_{\Omega}$ and $\gamma_g$ determined by the algebraic
consistency conditions of
\dm\ and the tadpoles of \GJ .  Note that there is an extra phase in 
the $\gamma$ matrices of \dm\ as compared to \GJ\ , i.e., $\tr
(\gamma^{-1}_{\Omega k,9}\gamma^T _{\Omega k,9})=e^{-2\pi ik/M}\tr
(\gamma_ {2k,9})$, etc.  The extra minus in the Klein bottle $\IZ _2$
twisted sector goes away, and we directly apply the tadpoles of
\GJ\ with $w\cdot n=32$ and
\eqn\gjnov{{1\over 4\sin^2 ({\pi k\over M})}(\sum_{\mu =0}^{M-1}
w_{\mu}e^{{2\pi i
k\mu\over M}}-4\sin^2 ({\pi k\over M})v_{\mu}e^{{2\pi i
k\mu\over M}}- 16\delta_{k,0\, mod\, 2}(1+e^{4\pi ik/M}))^2=0.}
Again these results give \anov\ after a brief calculation.  Note
that for $M$ odd, there are other theories with $0$ nine-branes, one of
which was discussed by \GJ .  In this case one expects two possible
projections of $\Omega$, one of which gives twisted strings.

\newsec{Interpretation of the extra tensor multiplets}

The above, perturbative, orientifold analysis applies far out along
the ``Coulomb branch,'' where the tensors have large expectation value
and the theory is weakly coupled.  Let us try to obtain an intuitive
picture for how the extra tensors of the Coulomb branch arise in type
I orientifolds.  In the context of type IIB at the ALE singularity,
there is a moduli space with $r$ ${\cal N}=(2,0)$ matter multiplets,
each consisting of a ${\cal N}=(1,0)$ tensor multiplet plus
hypermultiplet, giving $5r$ real scalars.  The $4r$ scalars arising
{}from the ${\cal N}=(1,0)$ hypermultiplets have expectation values
determined by the $3r$ blowing up parameters
\blowup\ and $B$ fields \uBreln, while the expectation values of the
$r$ scalars from the ${\cal N}=(1,0)$ tensor multiplets are more
subtle: as in \refs{\wcomments, \sopen, \witfb}, they can be
interpreted geometrically as the separations of five-branes in the
extra dimension of $M$ theory.  There are strings, arising from
wrapping three-branes on the cycles $\Sigma _i$, which couple to the
tensor multiplet and become tensionless where the tensor multiplet has
zero expectation value.

Upon orientifolding to obtain type I, one either keeps the ${\cal
N}=(1,0)$ hypermultiplet part of the ${\cal N}=(2,0)$ matter multiplet
or one keeps the ${\cal N}=(1,0)$ tensor multiplet.  We expect,
following the above discussion, that the expectation values of the
tensor multiplets which we obtained should have an interpretation as
distances of five-branes in the extra direction of $M$ theory.
Consider, for example $M$ theory on $\IR ^1/\IZ _2$.  Arguments
similar to \horwit, suggest that the theory at the origin of $\IR
^1/\IZ _2$ is the infinite coupling limit of the type I $SO(32)$
ten-dimensional string theory.  By ``compactifying'' this theory on
the infinite volume ALE space, we can obtain a finitely coupled type I
theory.  (We would need another dimension, as in $F$ theory, to obtain
a finitely coupled type I theory in ten dimensions.)  In analogy with
the $E_8$ case, we expect that a small $SO(32)$ instanton turns into
an NS five-brane which can wander away from the origin into $\IR /\IZ
_2$, with distance equal to the expectation value of the tensor
multiplet.

\bigskip
\centerline{{\bf Acknowledgments}}

  The work
of K.I. is supported by NSF PHY-9513835, the W.M. Keck Foundation, an
Alfred Sloan Foundation Fellowship, and the generosity of Martin and
Helen Chooljian.
\listrefs

\end